\documentclass{pasj00}

\SetRunningHead{Hyodo et al.}{Suzaku Spectroscopy of M17}

\begin{document}
\title{Suzaku Spectroscopy of the Extended X-Ray Emission in M17}
\author{Yoshiaki~\textsc{Hyodo},\altaffilmark{1}
Masahiro~\textsc{Tsujimoto},\altaffilmark{2,3,4}
Kenji~\textsc{Hamaguchi},\altaffilmark{5,6} Katsuji~\textsc{Koyama},\altaffilmark{1}\\
Shunji~\textsc{Kitamoto},\altaffilmark{2} Yoshitomo~\textsc{Maeda},\altaffilmark{7}
Yohko~\textsc{Tsuboi}\altaffilmark{8}, and Yuichiro~\textsc{Ezoe}\altaffilmark{9}}
\altaffiltext{1}{Department of Physics, Graduate School of Science, Kyoto University,\\
Kita-shirakawa Oiwake-cho, Sakyo, Kyoto 606-8502}
\altaffiltext{2}{Department of Physics, Rikkyo University, 3-34-1 Nishi-Ikebukuro, Toshima, Tokyo 171-8501}
\altaffiltext{3}{Department of Astronomy \& Astrophysics, Pennsylvania State University,\\
525 Davey Laboratory, University Park, PA 16802, USA}
\altaffiltext{4}{Chandra Fellow}
\altaffiltext{5}{CRESST and X-ray Astrophysics Laboratory, Goddard Space Flight Center,\\
National Aeronautics and Space Science, Greenbelt, MD 20771, USA}
\altaffiltext{6}{Universities Space Research Association, 10211 Wincopin Circle, Suite 500, Columbia, MD 21044, USA}
\altaffiltext{7}{Institute of Space and Astronautical Science, Japan Aerospace Exploration Agency,\\
3-1-1 Yoshinodai, Sagamihara, Kanagawa 229-8510}
\altaffiltext{8}{Department of Science and Engineering, Chuo University, 1-13-27 Kasuga, Bunkyo, Tokyo 112-8551}
\altaffiltext{9}{Cosmic Radiation Laboratory, RIKEN, 2-1 Hirosawa, Wako, Saitama 351-0198}

\email{hyodo@cr.scphys.kyoto-u.ac.jp}
\maketitle

\begin{abstract}
 We present the results of a Suzaku spectroscopic study of the soft extended X-ray
 emission in the H\emissiontype{II} region M17. The spectrum of the extended emission
 was obtained with a high signal-to-noise ratio in a spatially-resolved manner using the
 X-ray Imaging Spectrometer (XIS). We established that the contamination by unresolved
 point sources, the Galactic Ridge X-ray emission, the cosmic X-ray background, and the
 local hot bubble emission is negligible in the background-subtracted XIS spectrum of
 the diffuse emission. Half a dozen of emission lines were resolved clearly for the
 first time, including K$\alpha$ lines of highly ionized O, Ne, and Mg as well as L
 series complex of Fe at 0.5--1.5~keV. Based on the diagnosis of these lines, we
 obtained the following results: (1) the extended emission is an optically-thin thermal
 plasma represented well by a single temperature of $\sim$3.0\,$\pm$\,0.4~MK, (2) the
 abundances of elements with emission lines in the diffuse spectrum are 0.1--0.3~solar,
 while those of bright discrete sources are 0.3--1.5~solar, (3) the metal abundances
 relative to each other in the diffuse emission are consistent with solar except for a
 Ne enhancement of a factor of $\sim$2, (4) both the plasma temperature and the chemical
 composition of the diffuse emission show no spatial variation across the studied
 spatial scale of $\sim$5~pc.
\end{abstract}

\section{Introduction}
Massive stars are a driving force of physical and chemical evolutions of their host
galaxies. Supernova explosions and their remnants have been intensively studied for
decades, but the pre-explosion effects are equally important; the integrated mass,
momentum, and energy releases over the lifetime of an O star can be comparable to those
by a supernova explosion at the end of their lives \citep{leitherer92}. Diffuse X-ray
emission is generated as a consequence of shocks by stellar winds impinging on the
interstellar medium (ISM). Therefore, we can quantitatively study the effects of energy
dissipation and the chemical enrichment of interstellar space by early-type stars
through the spectroscopy of X-ray emission in H\emissiontype{II} regions.

\citet{weaver77} presented a self-similar solution of stellar winds interacting with the
ISM. They showed a single O7 star forms a hot ($\sim$10$^{6-7}$~K) bubble by the shock,
which can be observed as soft extended ($\sim$10~pc) X-ray emission. \citet{townsley03}
claimed the first unambiguous detections of such emission in the H\emissiontype{II}
regions M17 and the Rosette Nebula using the Advanced CCD Imaging Spectrometer (ACIS;
\cite{garmire03}) onboard the Chandra X-ray Observatory \citep{weisskopf02}. With a
$\sim$40~ks integration time of M17 (ObsID\,$=$\,972), diffuse soft X-ray emission was
detected apart from 886 point sources above $\sim$10$^{29.3}$~erg~s$^{-1}$
\citep{broos07}. \citet{dunne03} showed the entire structure of the soft X-ray diffuse
emission using the Position-Sensitive Proportional Counter (PSPC; \cite{pfeffermann87})
onboard ROSAT \citep{trumper82}. They measured the total X-ray luminosity and compared
to the wind-blown bubble models with and without heat conduction. They concluded that
only the bubble without heat conduction can account for the observed X-ray
luminosity of $\sim 2.5\times10^{33}$~erg~s$^{-1}$. The magnetic field may be
responsible for suppressing the heat conduction and mass evaporation between the hot gas
and cold ISM \citep{dunne03}.

The results obtained by these high-resolution imaging studies are generally consistent
with the wind-blown bubble models. However, the observational results are not still
accurate enough to compare to the theoretical works.  It is often ambiguous whether the
observed diffuse emission is from wind-blown bubbles or from supernovae, which give rise
to diffuse emission with a similar X-ray spectral hardness and luminosity in a similar
spatial scale. The largest uncertainty stems from the lack of spectral analysis based on
line diagnostics in a spatially-resolved manner.

Resolving emission lines is crucial to examine whether the spectrum is thermal and to
determine the temperature and the chemical composition of the plasma. For example, the
intensity ratio of K$\alpha$ lines between O\emissiontype{VII} and O\emissiontype{VIII}
and that between Ne\emissiontype{IX} and Ne\emissiontype{X} are steep functions of the
plasma temperature at 1--10~MK \citep{tucker66}. In between the O and Ne K$\alpha$
complex, Fe L series lines dominate the spectrum. The metallicity of these elements is
one of the factors to determine the X-ray luminosity expected from a wind-blown bubble
\citep{chu95}. The anomaly in the O and Fe abundance ratio can be used to discriminate
different types of supernovae \citep{tsujimoto95,nomoto97} if the emission is of a
supernova origin. The previous studies using ROSAT and Chandra were incapable of
resolving these lines, limiting their ability to diagnose the plasma emission.

The X-ray Imaging Spectrometer (XIS; \cite{koyama07}) onboard Suzaku \citep{mitsuda07}
has a superior spectral resolution, a low background, and a large effective area, which
are particularly suited for spectroscopy of extended X-ray emission. The capability to
resolve key elements with sufficient statistical significance has been illustrated by
several initial studies on extended emission in H\emissiontype{II} regions
\citep{hamaguchi07,tsujimoto07}. \citet{hamaguchi07} resolved various emission lines
from the diffuse emission in the Carina Nebula. Based on the low nitrogen-to-oxygen
ratio and the spatial variation of the Fe and Si abundances, they suggested that the
diffuse emission originates not from wind-blown bubbles but from one or multiple old
supernova remnant(s).

\medskip

M17 is a Galactic H\emissiontype{II} region at a distance of $\sim$1.6~kpc
\citep{nielbock01}. \citet{hanson97} identified nine O stars in the central OB
association with near infrared spectroscopy. A much larger number of young OB stars are
suggested by near infrared photometry \citep{lada91,jiang02}. The earliest system is a
binary of two O4--O5 stars. The age of the cluster is estimated to be $\lesssim$1~Myr
based on the H-R diagram \citep{hanson97}. Strong winds and radiation from the central
OB association sculpted the ambient matter to form a \textsf{V}-shaped cloud, which was
traced by molecular and atomic hydrogen lines
\citep{chrysostomou02,felli84,brogan01}. The diffuse X-ray emission found by Chandra and
ROSAT has an asymmetric morphology with respect to the OB association and fills the
cavity of the molecular cloud toward the negative Galactic latitude \citep{povich07}.

M17 is suitable for X-ray studies of hot bubbles in H\emissiontype{II} regions for being
proximate and having a very high contrast of the diffuse emission against point
sources. From the Chandra study \citep{townsley03}, the diffuse emission is more intense
than the integrated emission of resolved point sources by more than ten-fold at
1~keV. Despite the limited spatial resolution of Suzaku, therefore, the obtained
spectrum is not seriously contaminated by unresolved point sources.

\section{Observation}
Suzaku observed M17 on 2006 March 11--14 in the first announcement of opportunity
observing cycle. Suzaku has XIS and Hard X-ray Detector (HXD;
\cite{kokubun07,takahashi07}) instruments. We concentrate on the XIS data in this paper,
which has a sensitivity for the soft emission studied here.

The XIS is equipped with four X-ray CCDs. Three of them (XIS0, 2, and 3) are
front-illuminated (FI) CCDs and the remaining one (XIS1) is a back-illuminated (BI)
CCD. Each CCD chip has a format of 1024$\times$1024 pixels and is composed of four
segments of 256$\times$1024 pixels. FI and BI CCDs are superior to each other in the
hard and soft band responses, respectively. They are mounted at the focus of four
independent X-ray telescopes (XRT; \cite{serlemitsos07}). The detectors are sensitive in
the energy range of 0.2--12.0~keV with an initial energy resolution of $\sim$65~eV in
the full width at half maximum and a total effective area of $\sim$1360~cm$^{2}$ at
1.5~keV. An XIS field of view covers a $\sim$18\arcmin $\times$18\arcmin\ region with a
half power diameter of $\sim$2\arcmin. The radioactive sources of \atom{Fe}{}{55}
illuminate two corners of each of the four CCDs for calibration purposes.

The performance of XIS is subject to degradation due to the radiation damage in the
orbit \citep{koyama07}. As of the observation date, the energy resolution in the full
width at half maximum is $\sim$90~eV at 1.5~keV. The relative energy gains among XIS
chips and segments have a systematic uncertainty of $\sim$5~eV. An unknown contaminant
accumulates on the optical blocking filters of the XIS in the orbit. As a result, the
effective area at the soft band end has diminished significantly, making the carbon and
nitrogen features difficult to detect below $\sim$0.5~keV. This effect is included in
the auxiliary response.

The observation was conducted using the normal clocking mode with a frame time of
8~s. Data (revision 1.2\footnote{See http://www.astro.isas.jaxa.jp/suzaku/process/ for
details.}) were screened to remove events during the South Atlantic Anomaly passages, at
elevation angles below 4$^\circ$ from the earth rim, and at elevation angles below
10$^\circ$ from sunlit earth rim. We constructed plots of the raw count rate versus the
elevation angles to find that these criteria maximize the exposure time with negligible
contaminating emission. After the filtering, the net integration time is $\sim$110~ks.

\section{Analysis}
\subsection{Image Analysis}
\begin{figure*}
 \begin{center}
  \FigureFile(85mm,85mm){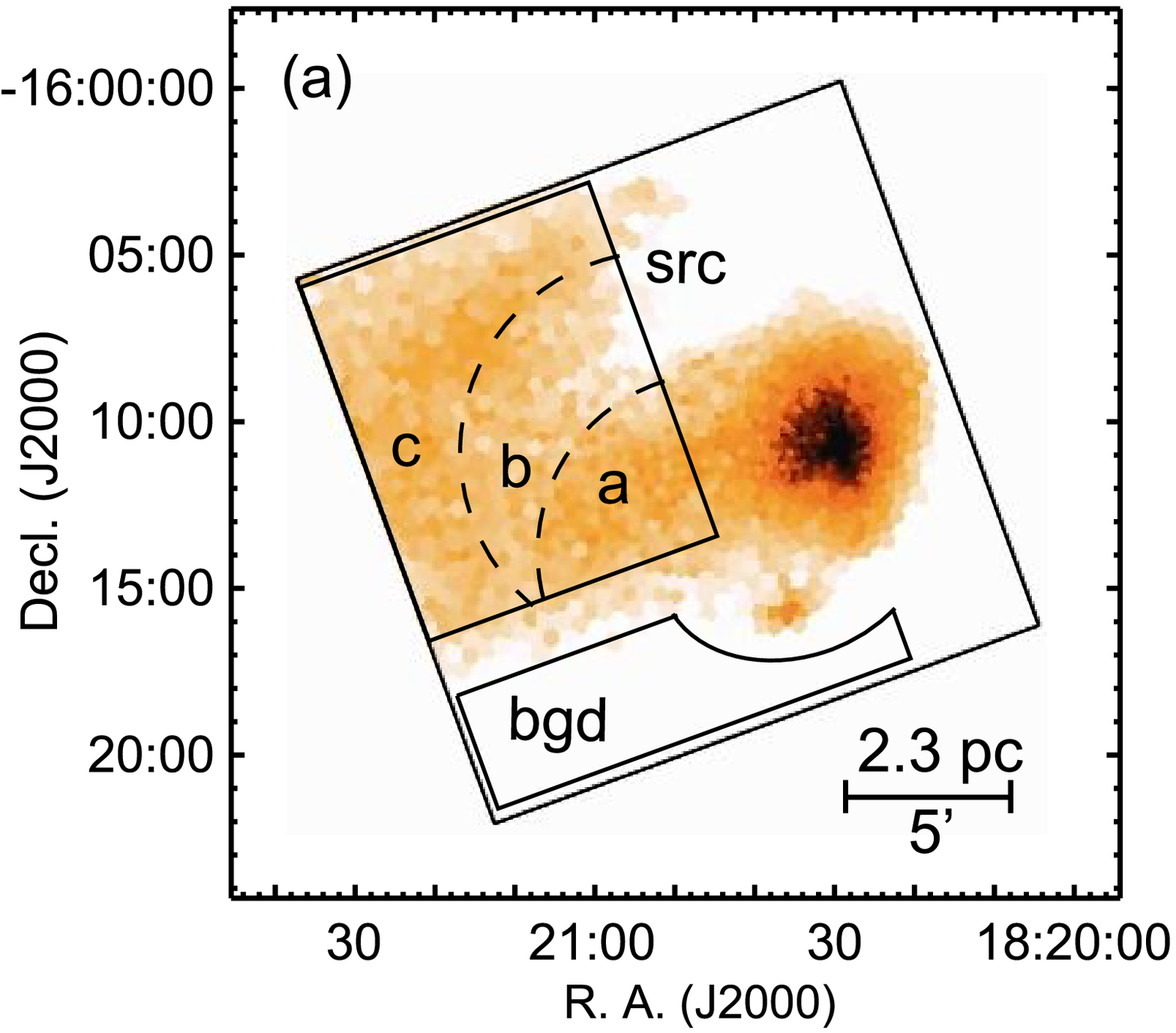}
  \FigureFile(85mm,85mm){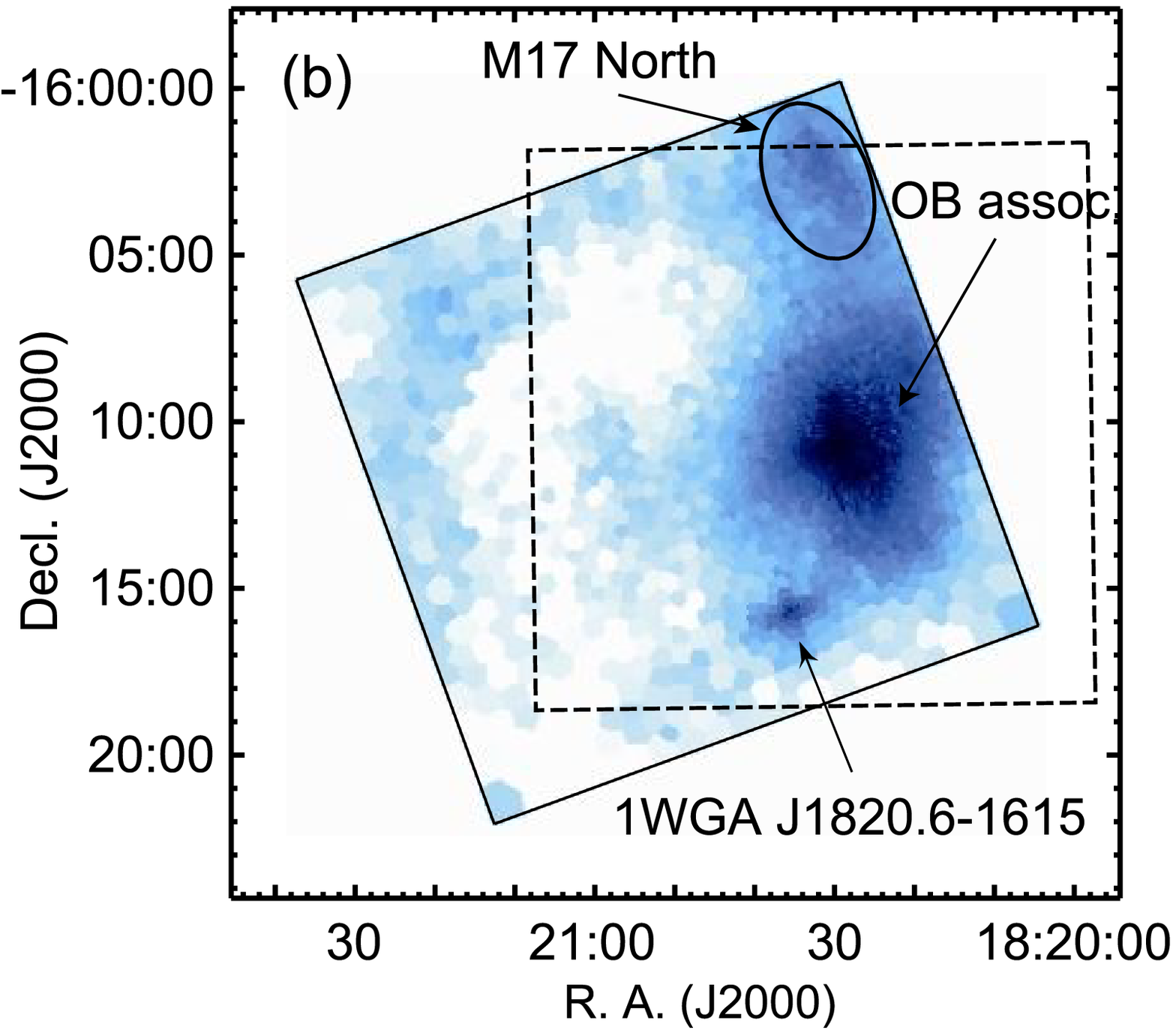}
 \end{center}
 \caption{XIS images in the (a) 0.5--1.5~keV and (b) 1.5--5~keV bands. The XIS field is
 shown with the solid square in both panels, while the ACIS field is shown with the
 dashed square in (b). The source and background regions for diffuse emission are shown
 by solid lines in (a). The source sub-regions (a, b, and c) are ruled by dashed
 curves. The discrete sources are shown with arrows in (b). Both images are processed as
 follows: (1) Non--X-ray background constructed from night earth observations was
 subtracted. (2) The astrometry and vignetting were corrected. (3) The images were
 adaptively binned to achieve a signal-to-noise ratio of larger than 8 using the
 weighted Voronoi tessellation algorithm (\cite{Diehl06},
 \cite{cappellari03}).}\label{fg:f1}
\end{figure*}

Figure~\ref{fg:f1} shows the XIS images of the study field in the (a) soft
(0.5--1.5~keV) and (b) hard (1.5--5.0~keV) bands. The two band-limited images appear
strikingly different. In the hard band, the image is dominated by the emission from the
OB association. We also see the excess emission from a group of protostars M17 North
\citep{wilson79,henning98,broos07}. In the soft band, the extended emission emerges in
the eastward of the OB association, as was claimed by \citet{townsley03} and
\citet{dunne03}. The XIS observation was centered at the most intense part of the
extended emission at (R.A., decl.) $\sim$ (\timeform{18h20m50s}, --\timeform{16D12'}) in
the equinox J2000.0, while the ACIS observation was at the OB association
\citep{townsley03}. The XIS and ACIS images have a similar size ($\sim$18 and $\sim$17
arc-minute square, respectively) with a $\sim$70\% overlapping area
(figure~\ref{fg:f1}b).

The astrometry of the XIS frame was registered using 1WGA\,J1820.6--01615 found in both
images. The O8 star \citep{ogura76,white94} is bright, isolated, and point-like in the
ACIS image (CXOU\,J182035.87--161542.5; \cite{broos07}), thus serves as a good
astrometric calibrator. We shifted the XIS frame by $\sim$17\arcsec\ to the north so
that the position matches with that by the ACIS observation. The Chandra frame is
accurate to $\sim$0\farcs5 in the astrometry\footnote{See
http://cxc.harvard.edu/proposer/POG for details.}.

\subsection{Spectral Analysis of Extended Emission}
\subsubsection{Entire Emission}
We first examine the spectrum of the entire diffuse emission. We extracted the source
spectrum from a rectangular region and the background spectrum from a region devoid of
intense diffuse emission (figure~\ref{fg:f1}a). Because the off-axis angles of the
source and background regions are different, we processed the raw spectra in the
following way before subtracting the background from the source: (1) The
non--X-ray-background (NXB) spectrum was subtracted, which was constructed from night
earth data at the same extraction region. The NXB of XIS is a function of the
geomagnetic cut-off rigidity. We therefore compiled night earth observations such that the
cut-off rigidity distribution becomes the same with that of the M17 observation. (2) The
vignetting was corrected by multiplying the effective area ratios between the source and
background regions for each energy bin of the background spectrum. This takes into
account the accumulating contaminant on the XIS optical blocking filter.

\begin{figure}[!ht]
 \begin{center}
  \FigureFile(85mm,85mm){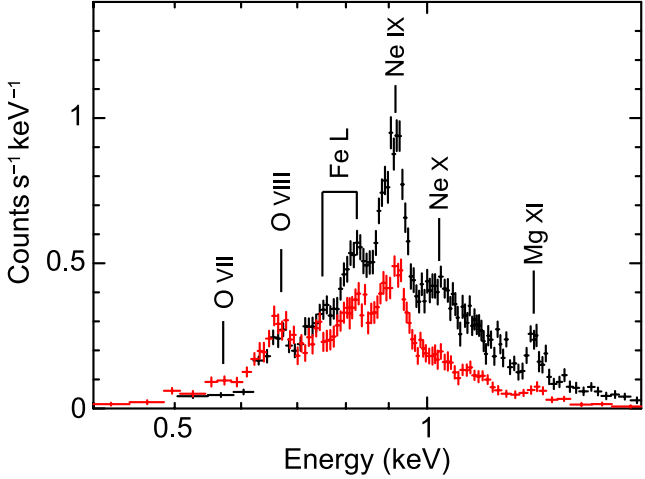}
 \end{center}
 \caption{Background-subtracted XIS spectra of the entire diffuse emission (regions
 a$+$b$+$c in figure~\ref{fg:f1}a). BI spectrum is shown in red, while the merged FI
 spectrum is in black in a linear scale. Conspicuous emission lines are labeled, which are
 K$\alpha$ lines except for the Fe L series lines.}\label{fg:f2}
 \end{figure}

The merged FI and the BI spectra are shown in figure \ref{fg:f2}. In the merged FI spectrum,
we added the three FI spectra to increase the photon statistics, because the
redistribution matrix functions (RMFs) and the auxiliary response functions (ARFs) are
essentially the same for these chips. On the other hand, we handled the BI spectrum
separately for its different response.

We resolved the emission lines clearly for the first time, which include K shell lines
of O, Ne, and Mg as well as L shell lines of Fe. This indicates that the emission is of
a thermal origin. The O\emissiontype{VIII} line is much stronger than the
O\emissiontype{VII} line and the Ne\emissiontype{IX} line is so than the
Ne\emissiontype{X} line if we taking the energy dependence of the efficiencies into
account. These line ratios alone infer that the plasma temperature is in the range of
2.5--4~MK \citep{tucker66} even without spectral model fittings.

\subsubsection{Spatially-Resolved Emission}
\begin{figure}
 \begin{center}
  \FigureFile(85mm,240mm){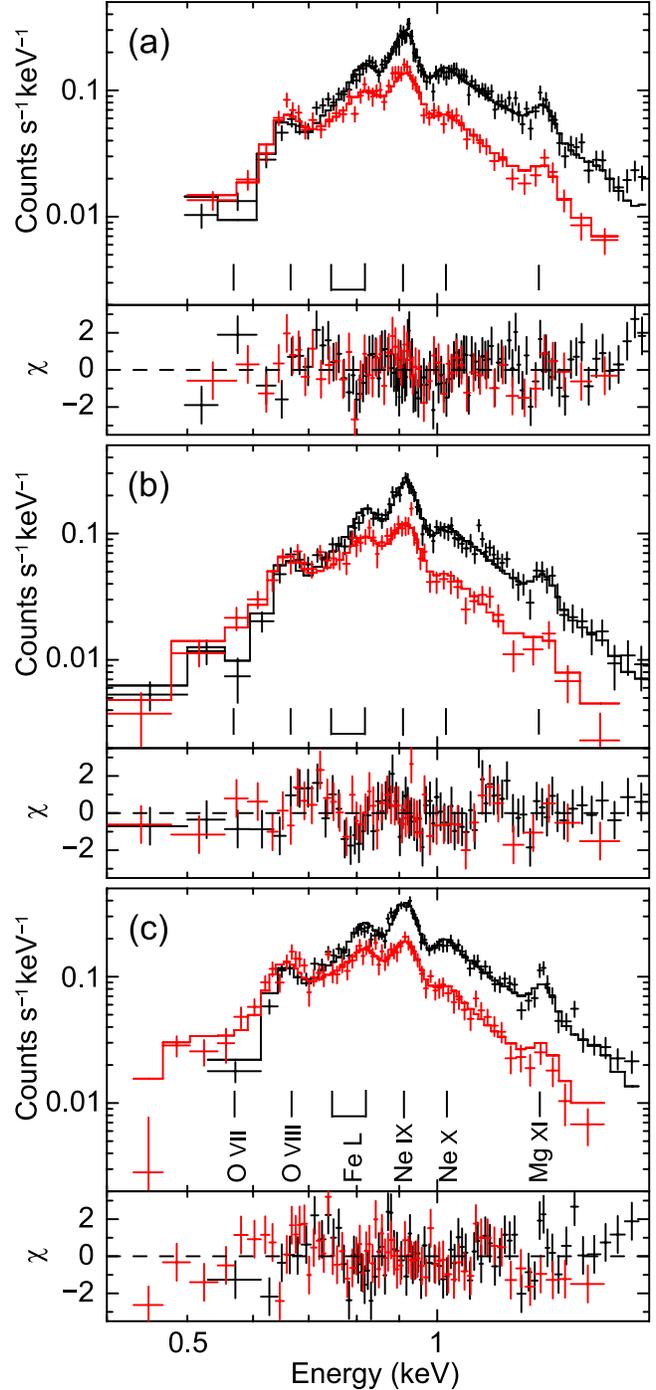}
 \end{center}
 \caption{Background-subtracted 0.4--1.8~keV band spectra of the diffuse emission in the
 three regions (a), (b), and (c) in figure \ref{fg:f1} (a). The BI spectrum is shown in red,
 while the merged FI spectrum is in black in a logarithmic scale. The upper panels show
 the data in crosses and the best-fit models in solid lines, while the lower panels show
 residuals to the fit.}\label{fg:f3}
\end{figure}

In order to investigate the spatial difference of the plasma properties, we divided the
source region into three sub-regions (a, b, and c in figure \ref{fg:f1}a) based on the
morphology of the diffuse emission. We constructed the spectra from each region and 
subtracted the background in the same manner for the entire emission. The merged FI and
the BI spectra in each sub-region are shown in figure~\ref{fg:f3}.

We fitted the 0.4--1.8 keV spectra with a thin-thermal plasma model at a collisional
equilibrium (the APEC model; \cite{smith01}) convolved with the interstellar absorption
\citep{morrison83}. The abundances of the noticeable elements (O, Ne, Mg, and Fe) were
free parameters. Those of the other elements were fixed at 0.3 solar, which is
canonically used in X-ray spectroscopy in star-forming regions (e.g.,
\cite{getman05}). We used the RMFs (version 2006-08-01) of the observation month and
generated ARFs using a ray-tracing simulator (\texttt{xissimarfgen} version 2006-08-28;
\cite{ishisaki07}) assuming that the emission is uniform across a 15\arcmin\ radius
circle centered at the optical axis. In order to compensate for the possible uncertainty
in the energy gain calibration, we introduced an additional fitting parameter
(offset). The resultant offset values were 1--4 eV, which are within the current
calibration limitation.

A single temperature model yielded acceptable fits for all spectra. The best-fit
hydrogen-equivalent column density ($N_{\rm H}$), plasma temperature ($k_{\rm{B}}T$),
metallicity ($Z_{\rm{O}}$, $Z_{\rm{Ne}}$, $Z_{\rm{Mg}}$, and $Z_{\rm{Fe}}$), the average
surface brightness ($S_{\rm{X}}$), and the luminosity ($L_{\rm{X}}$) in the 0.5--2.0~keV
band are summarized in table~\ref{tb:t1}. The best-fit models are shown in
figure~\ref{fg:f3}. We attempted different plasma models with multiple temperatures or\
non-equilibrium ionization, but did not obtain improved fits. We therefore consider that
a single temperature model at a collisional equilibrium is adequate.

\subsection{Spectral Analysis of Discrete Sources}
\begin{figure}
 \begin{center}
  \FigureFile(85mm,240mm){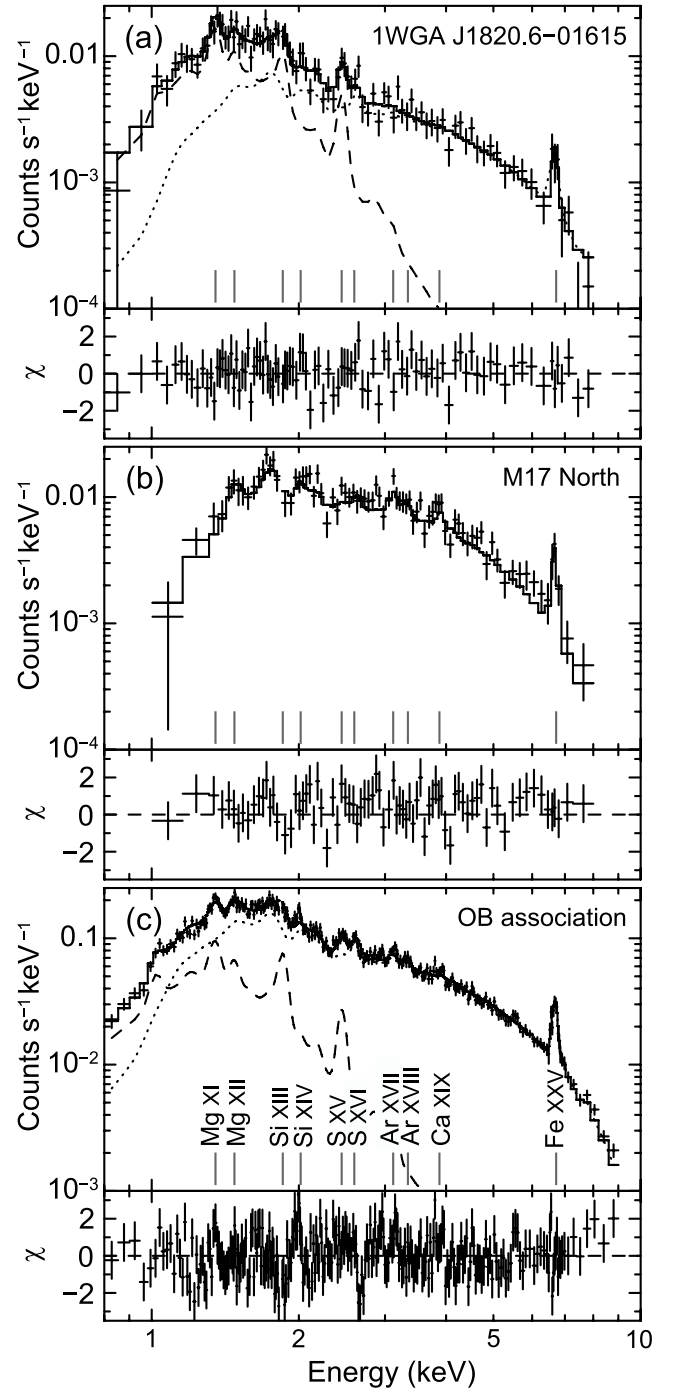}
 \end{center}
 \caption{Background-subtracted 0.8--10~keV band spectra of the three discrete sources:
 (a) 1WGA\,J1820.6--1615, (b) M17 North, and (c) the OB association. The symbols follow
 figure~\ref{fg:f3}. The higher and lower temperature components of the two temperature
 model are shown respectively with dashed and dotted lines. For simplicity, only the
 merged FI spectra and the best-fit models are shown.}\label{fg:f4}
\end{figure}

We also constructed spectra of the three discrete sources (1WGA\,J1820.6--1615, M17
North, and the OB association in figure~\ref{fg:f1}b). The source signals were accumulated
from elliptical regions of 2\farcm0--3\farcm5 axis lengths, while the background signals
were from adjacent regions free of sources. The spectra are shown in
figure~\ref{fg:f4}. All the spectra are characterized by hard emission full of K$\alpha$
emission lines of highly ionized ions, which include Mg\emissiontype{XI},
Mg\emissiontype{XII}, Si\emissiontype{XIII}, Si\emissiontype{XIV}, S\emissiontype{XV},
S\emissiontype{XVI}, Ar\emissiontype{XVII}, Ar\emissiontype{XVIII},
Ca\emissiontype{XIX}, and Fe\emissiontype{XXV}.

We can have a crude estimate for spectral models from the intensity of these lines. The
spectra of 1WGA~J1820.6--1615 and the OB association show a Mg\emissiontype{XI} line
stronger than Mg\emissiontype{XII} as well as Fe\emissiontype{XXV} and a hard continuum
up to $\sim$8~keV, requiring at least two thermal components of different
temperatures. M17 North shows a Mg\emissiontype{XII} line and no prominent
Mg\emissiontype{XI} line, thus a single temperature model would be adequate.

We fitted these spectra using the attenuated thin-thermal plasma model similarly for the
diffuse emission. The merged FI and the BI spectra were simultaneously fitted for
1WGA\,J1820.6--1615 and the OB association. Only the merged FI spectrum was fitted for
the M17 North because its BI spectrum is strongly contaminated by the \atom{Fe}{}{55}
calibration source. We first fitted the spectra with a one-temperature model. An
additional component with a different temperature was added if the fitting was rejected
due to systematic residuals. As the crude estimate, the spectra of 1WGA\,J1820.6--1615
and the OB association required a two-temperature model, while that of the M17 North was
fitted by a one-temperature model. The best-fit models and parameters are shown in
figure~\ref{fg:f4} and table~\ref{tb:t2}, respectively.

\begin{table*}
 \begin{center}
  \caption{Best-fit APEC parameters for the spatially-resolved diffuse spectra.}\label{tb:t1}
  \begin{tabular}{llccc}
   \hline
   \hline
   Pars. & Units & \multicolumn{3}{c}{\hrulefill\hspace*{5mm}Sub-regions\hspace*{5mm}\hrulefill} \\
              &       & (a) & (b) & (c) \\
   \hline
   $N_{\rm H}$\footnotemark[$*$]   & (10$^{21}$~cm$^{-2}$) \dotfill 
       & 4.8 (4.4--5.2) & 4.6 (4.4--5.2) & 4.3 (3.7--4.6) \\
   $k_{\rm{B}}T$\footnotemark[$*$] & (keV)   \dotfill 
       & 0.25 (0.24--0.27) & 0.24 (0.22--0.25) & 0.27 (0.25--0.28) \\
   $Z_{\rm{O}}$\footnotemark[$*$]  & (solar) \dotfill 
       & 0.10 (0.06--0.14) & 0.15 (0.11--0.18) & 0.13 (0.10--0.16) \\
   $Z_{\rm{Ne}}$\footnotemark[$*$] & (solar) \dotfill 
       & 0.20 (0.17--0.24) & 0.32 (0.25--0.39) & 0.22 (0.19--0.26) \\
   $Z_{\rm{Mg}}$\footnotemark[$*$] & (solar) \dotfill 
       & 0.10 (0.09--0.13) & 0.12 (0.07--0.16) & 0.12 (0.09--0.16) \\
   $Z_{\rm{Fe}}$\footnotemark[$*$] & (solar) \dotfill 
       & 0.10 (0.08--0.13) & 0.19 (0.14--0.20) & 0.12 (0.11--0.14) \\
   $S_{\rm{X}}$\footnotemark[$*\dagger$] & (10$^{-14}$ erg~s$^{-1}$~cm$^{-2}$~arcmin$^{-2})$ \dotfill 
       & 2.16 (2.12--2.21) & 1.34 (1.31--1.38) & 1.55 (1.52--1.58) \\
   $L_{\rm{X}}$\footnotemark[$\ddagger$] & (10$^{33}$~erg~s$^{-1}$) \dotfill 
       & 1.3 & 1.0 & 1.2\\
   $\chi^{2}$/d.o.f & \dotfill
       & 155.5/146 & 121.8/114 & 176.9/129 \\ 
   \hline
   \multicolumn{5}{@{}l@{}}{\hbox to 0pt{\parbox{170mm}{\footnotesize
   \par\noindent
   \footnotemark[$*$] The uncertainties in the parentheses are the 90\% confidence range.
   \par\noindent
   \footnotemark[$\dagger$] The average X-ray surface brightness in the 0.5--2.0~keV band. 
   \par\noindent
   \footnotemark[$\ddagger$] The absorption-corrected X-ray luminosity in the 0.5--2.0~keV band. A distance of 1.6~kpc is assumed.
   }\hss}}
  \end{tabular}
 \end{center}
\end{table*}

\begin{table*}
 \begin{center}
  \caption{Best-fit APEC parameters for the discrete spectra.}\label{tb:t2}
  \begin{tabular}{llccc}
   \hline
   \hline
   Pars. & Units & 1WGA\,J1820.6--01615 & M17 North & OB association \\
   \hline
   $N_{\rm H}$\footnotemark[$*$] & (10$^{22}$~cm$^{-2}$) \dotfill 
       & 1.7 (1.4--2.0) & 2.4 (2.1--2.7) & 1.3 (1.2--1.4) \\
   $k_{\rm{B}}T_{\rm high}$\footnotemark[$*\dagger$] & (keV) \dotfill
       & 3.8 (3.1--4.7) &2.7 (2.4--3.2) & 4.0 (3.9--4.1) \\
   $k_{\rm{B}}T_{\rm low}$\footnotemark[$*\dagger$] & (keV) \dotfill 
       & 0.56 (0.47--0.65) & ... &0.59 (0.56--0.62) \\
   $Z_{\rm{Ne}}$\footnotemark[$*$] & (solar) \dotfill 
       & 0.3 &0.3 & 0.51 (0.34--0.71) \\
   $Z_{\rm{Mg}}$\footnotemark[$*$] & (solar) \dotfill
       & 0.49 (0.24--0.87) & 0.3 & 0.51 (0.38--0.65) \\
   $Z_{\rm{Si}}$\footnotemark[$*$] & (solar) \dotfill
       & 0.30 (0.16--0.58) & 0.3 & 0.53 (0.39--0.71) \\
   $Z_{\rm{S}} $\footnotemark[$*$] & (solar) \dotfill 
       & 1.07 (0.58--1.72) & 0.3 & 1.17 (0.98--1.38) \\
   $Z_{\rm{Ar}}$\footnotemark[$*$] & (solar) \dotfill
       & 0.3 & 1.67 (0.72--3.00) & 1.44 (1.00--1.90) \\
   $Z_{\rm{Ca}}$\footnotemark[$*$] & (solar) \dotfill
       & 0.3 & 0.3 & 0.46 (0.01--0.90) \\
   $Z_{\rm{Fe}}$\footnotemark[$*$] & (solar) \dotfill
       & 0.33 (0.19--0.48) & 0.36 (0.22--0.50) & 0.28 (0.25--0.31) \\
   $F_{\rm{X}}$\footnotemark[$*\ddagger$] & (10$^{-13}$ erg~s$^{-1}$~cm$^{-2}$) \dotfill
       & 3.9 (3.8--4.0) & 7.4 (7.1--7.7) & 58.3 (57.8--58.8) \\
   $L_{\rm{X}}$\footnotemark[$\S$]& (10$^{32}$~erg~s$^{-1}$) \dotfill
       & 3.2 & 4.9 & 30.0 \\ 
   $\chi^2$/d.o.f. & \dotfill 
       & 100.2/134 & 66.4/76 & 473.3/405 \\
   \hline
   \multicolumn{5}{@{}l@{}}{\hbox to 0pt{\parbox{150mm}{\footnotesize
   \par\noindent
   \footnotemark[$*$] The uncertainties in the parentheses are the 90\% confidence
   range. Fixed  values are shown without ranges.
   \par\noindent
   \footnotemark[$\dagger$] The plasma temperatures for the higher and lower temperature
   components. Only the higher temperature value is given for M17 North, which is fitted
   by a single temperature model.
   \par\noindent
   \footnotemark[$\ddagger$] The X-ray flux in the 1.0--8.0~keV band. 
   \par\noindent
   \footnotemark[$\S$] The absorption-corrected X-ray luminosity in the 1.0--8.0~keV band. A distance of 1.6~kpc is assumed.
   }\hss}}
  \end{tabular}
 \end{center}
\end{table*}

\section{Discussion}
\subsection{Contamination to the Extended Emission}\label{sect:s4-1}
The spectra of the diffuse emission in figures~\ref{fg:f2} and \ref{fg:f3} are
contaminated by other sources of emission. We evaluate the levels of contamination by
unresolved point sources, the Galactic Ridge X-ray emission (GRXE), the cosmic X-ray
background (CXB), and the local hot bubble (LHB), and argue that their contributions to
the background-subtracted spectra are negligible. About 70\% of the XIS field is covered
in the Chandra observation (figure~\ref{fg:f1}b), which has a much better spatial
resolution and sensitivity for faint point sources \citep{townsley03,broos07}. Among the
three sub-regions, the region (a) has a complete coverage by Chandra. We therefore use
this sub-region as a representative to evaluation the levels of various contaminations.

\begin{figure}
 \begin{center}
  \FigureFile(85mm,60mm){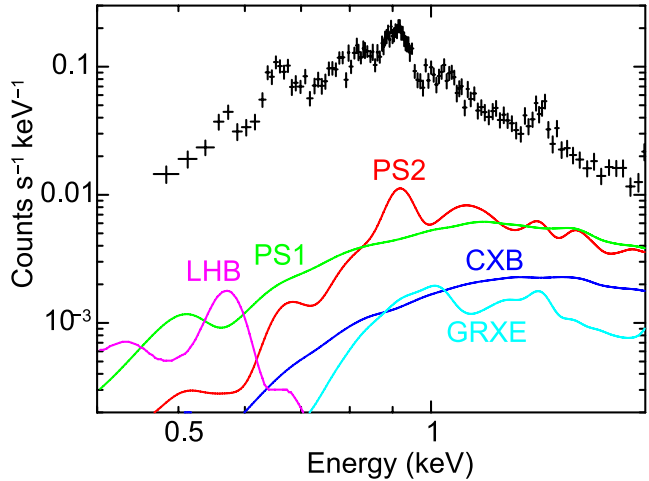}
 \end{center}
 \caption{Comparison of the observed diffuse spectrum to the simulated spectra of
 contaminating sources. The NXB signal is subtracted from the observed spectrum. PS1
 and PS2 are for the contribution of unresolved point sources with different spectral
 models, GRXE for the Galactic Ridge X-ray emission, CXB for the cosmic X-ray
 background, and LHB for the local hot bubble. None of them play a significant
 contribution. Moreover, most of them are removed by subtracting a background
 spectrum from the neighboring region.}\label{fg:f5}
\end{figure}

First, we examine the contribution of unresolved point sources. We extracted 19 point
sources from the ACIS data in the sub-region (a), constructed the composite spectrum,
and fitted it with a thermal plasma model. The spectrum of each source is too poor to
fit individually, so we assume that all sources have the same spectral shape that best
describes the composite spectrum. Using their positions, flux, and the assumed spectral
shape, we generated their XIS events using a ray-tracing simulator (\texttt{xissim};
\cite{ishisaki07}). In figure~\ref{fg:f5}, we compare the integral of the simulated
spectra of unresolved point sources (PS1) to the observed diffuse spectrum (pluses). In
the displayed observed spectrum, we subtracted the NXB spectrum but not the background
spectrum in the neighboring region. The point source contribution accounts for $\sim$8\%
of the emission in the 0.4--1.8~keV band.

To have an estimate of the maximum contamination to the lines, we repeated the same
procedure using the model with the maximum allowable abundance values in the best-fit
model for simulating unresolved point source events. The resultant integrated spectrum
(PS2) is also shown in figure~\ref{fg:f5}. At the Ne\emissiontype{IX} line at 0.92~keV,
the contribution by the point sources accounts only for $<$10\% of the observed
emission.

Second, we derive the GRXE contribution, which is ubiquitous along the Galactic Plane at
Galactic longitudes ($l$) of $|l| \lesssim$~45\arcdeg\ \citep{kaneda97,sugizaki01}. We
refer to a Chandra result \citep{ebisawa05} for the spectral shape in the soft band and
to a Rossi X-ray Timing Explorer result \citep{revnivtsev06} for the surface brightness
at the position of M17 ($l \sim$~15\arcdeg). The GRXE contribution to the observed
emission is estimated to be $\sim$3\% and is shown in figure~\ref{fg:f5}.

Third, for the CXB contribution, we consulted the Suzaku XIS observation of the North
Ecliptic Pole \citep{fujimoto07} both for the spectral shape and the surface
brightness. The CXB was observed and fitted by a power-law model. We convolved the model
with the XIS responses and found that the contribution is $\sim$2\%.

Finally, we constrain the contribution by the LHB emission. We estimate its surface
brightness to be $\sim$4$\times$10$^4$~counts~s$^{-1}$~arcmin$^{-2}$ in the PSPC R1 and
R2 bands \citep{snowden98}. We assume that the spectrum is a thin-thermal plasma (the
Raymond-Smith model; \cite{raymond77}) with a temperature of 0.1~keV. With the derived
emission measure of $\sim$2.8$\times10^{-3}$~cm$^{-6}$~pc, the LHB contributes $\sim$2\%
of the observed emission.

None of the above components play a significant role in the observed diffuse
spectrum. Moreover, we removed most of them by subtracting a background spectrum from a
neighboring region. The background events show no indication of time variability
caused by solar flares \citep{fujimoto07}. We conclude that the background-subtracted
XIS spectrum represents the spectrum of the uncontaminated diffuse emission quite well.

The background spectrum accounts for $\sim$35\% of the source spectrum of the sub-region
(a). Besides the NXB contributing $\sim$7\% of the source flux, the sum of unresolved
point sources, GRXE, CXB, and LHB contributes $\sim$15\%. Therefore, emission with a
flux of $\sim$13\% of the source flux is additionally included in the background
spectrum. The contribution from the bright sources outside of the extraction region is
negligible. We attribute the remaining emission to the diffuse emission in the
background region (figure~\ref{fg:f1}a). If this is the case, the flux estimate of the
diffuse emission (table~\ref{tb:t1}) is underestimated by $\sim$13\%.

\subsection{Comparison with Previous Studies}\label{sect:s4-2}
We compare our results with the previous works using ROSAT \citep{dunne03} and Chandra
\citep{townsley03,broos07}. We derived that the diffuse spectrum is explained by an
absorbed single temperature thin-thermal plasma model of $k_{\rm{B}}T \sim$~0.25~keV and
$N_{\rm H} \sim$~4.5$\times$10$^{21}$~cm$^{-2}$. The total luminosity (0.5--2.0~keV) in
the combined (a)$+$(b)$+$(c) region is 
$\sim$3.5$\times$10$^{33}$~erg~s$^{-1}$ (table~\ref{tb:t1}).

The ROSAT study \citep{dunne03} shows that the total luminosity of the diffuse emission
is $\sim$2.5$\times$10$^{33}$~erg~s$^{-1}$. The smaller estimate than our result is more
noticeable if we consider that the ROSAT value was derived from a larger area and in a
wider energy range (0.1--2.4~keV). This stems from an underestimate of the

extinction. Without a sufficient spectral resolution to resolve lines, two different
models were not disentangled in the ROSAT PSPC spectra; one is a low plasma temperature
with a large extinction ($\sim 0.2$~keV and $\sim 10^{22}$~cm$^{-2}$) and the other is a
high plasma temperature with a small extinction ($\sim$0.7~keV and $\sim
10^{20}-10^{21}$~cm$^{-2}$). \citet{dunne03} derived the luminosity based on the latter,
but our spectroscopy shows that the former should be in the case.

The best-fit XIS values of $L_{\rm{X}}$ and $N_{\rm H}$ are consistent with those
presented in the Chandra study \citep{townsley03}, in which
$L_{\rm{X}}=$~3.4$\times$10$^{33}$~erg~s$^{-1}$ (0.5--2.0~keV), and $N_{\rm
H}=$~(4~$\pm$~1)$\times$10$^{21}$~cm$^{-2}$. However, the plasma temperatures are
different between the two studies. In Chandra, the primary component of
$k_{\rm{B}}T=$~0.6~$\pm$~0.1~keV and the secondary component of $\sim$0.13~keV were
claimed. In Suzaku, however, we confirmed that a single temperature component of
$\sim$0.25~keV is adequate from the diagnosis of resolved oxygen and neon lines.

From the discrete sources, we detected K$\alpha$ emission lines from highly ionized ions
(figure~\ref{fg:f4}). The spectrum of the OB association is comprised of hundreds of
point sources, but the emission from an O4--O5 binary dominates the
spectrum. \citet{broos07} claimed that both components of the binary (sources 543 and
536) have plasma temperatures exceeding 10~keV. However, the strong Fe\emissiontype{XXV}
K$\alpha$ line at 6.7~keV and the weak Fe\emissiontype{XXVI} K$\alpha$ line at 7.0~keV
in the XIS spectrum (figure~\ref{fg:f4}c) do not support such high temperatures.

\subsection{Spatial Difference of the Plasma Properties}\label{sect:s4-3}
A high signal-to-noise ratio spectrum by XIS enabled us to conduct spatially-resolved
spectroscopy of the diffuse emission. The plasma temperature and the chemical
composition are uniform, except possibly for a larger metallicity in the sub-region
(b). The Chandra study \citep{townsley03} also show no evidence for spatial variation of
plasma temperature.

The observed uniformity indicates that the entire plasma is at a thermal equilibrium in
the observed spatial scale of $\sim$5~pc, unless the plasma is patchy at equilibria
locally by magnetic confinement. The global equilibrium is reasonable considering the
fact that the plasma sound crossing time ($\sim$2$\times$10$^{4}$~yr) is much smaller
than the time scale of the system ($\sim$10$^{6}$~yr), thus the constant pressure is
achieved \citep{weaver77}. Here, we used the plasma volume and the electron density
as $\sim$30~pc$^{3}$ and $\sim$1~cm$^{-3}$, respectively, by assuming that the plasma
distribution has a conical shape with its apex at the OB association and with a filling
factor of 1. Given the uniformity of the plasma temperature and pressure, we speculate
that the density is also spatially uniform.

The observed surface brightness, however, is different among the three sub-regions
(figure~\ref{fg:f1}a, table~\ref{tb:t1}). It is $\sim$1.6 times more intense in the
sub-region (a) than (b). This is not attributable entirely to the different extinction,
as $N_{\rm H}$ is larger in the sub-region (a) than in (b). Because the plasma has a
uniform temperature and density, we speculate that the difference of the surface
brightness is likely due to the different line-of-sight depths or different filling
factors.

With the derived plasma volume and the density, the total mass of the plasma is
$\sim$1~$M_{\odot}$. This is comparable to the integrated ejecta mass by stellar winds
at a mass loss rate of $\sim$10$^{-6}$~$M_{\odot}$ for $\sim$10$^{6}$~yr and agrees with
the estimates by the previous works \citep{dunne03,townsley03}. We also speculate that
the swept-up and evaporated ISM does not make a significant contribution to the plasma
mass.

\subsection{Chemical Composition}\label{sect:s4-4}
The chemical composition of the diffuse emission is revealed for the first time in this
study. The metallicity of the diffuse emission is 0.1--0.3 solar (table~\ref{tb:t1}),
which is significantly lower than those derived for the discrete sources (0.3--1.5
solar). The metal abundances relative to each other in the diffuse emission are
consistent with solar among O, Mg, and Fe \citep{anders89}, but Ne is enhanced in all
three sub-regions by a factor of $\sim$2 (table~\ref{tb:t1}). This is also evident in
the simultaneous spectral fits of the three sub-regions, in which we tied the abundance
values of these elements. The resultant values are $Z_{\rm{O}}=$~0.12 (0.11--0.13),
$Z_{\rm{Ne}}=$~0.22 (0.21--0.25), $Z_{\rm{Mg}}=$~0.10 (0.08--0.12), and
$Z_{\rm{Fe}}=$~0.12 (0.11--0.13) solar.

Such Ne enhancement from other metals is widely seen in coronally active stars
\citep{brinkman01,kastner02,audard03,imanishi03,stelzer04,maggio07} for unknown
reasons. We consider that the Ne enhancement is an intrinsic feature of the diffuse
plasma, and is not influenced by the contamination of Ne-enhanced point sources spectra
(\S~\ref{sect:s4-1}). One explanation for the anomaly is that the poorly-constrained
solar Ne abundance is underestimated by a factor of a few \citep{drake05,liefke06}. This
would account for the observed Ne enhancement in the M17 diffuse plasma as well.

The lack of a clear spatial variation of the chemical composition comprises a sharp
contrast to the diffuse emission in the Carina Nebula \citep{hamaguchi07}, where the
different abundance patterns across a similar spatial scale suggest the supernova origin
for the emission. In the diffuse emission in M17, the O and Fe ratio is consistent with
the solar abundance, which is another line of evidence against the supernova
interpretation. If a supernova has occurred in M17, it should have been caused by a
star earlier than the earliest (O4--O5) star in the OB association. Such a massive
source causes a core-collapse--type supernova. It would have yielded a measurably larger
ratio of O against Fe than the solar value by a factor of a few
\citep{tsujimoto95,nomoto97}.

\section{Summary}
We conducted a spectroscopic study of the soft diffuse X-ray emission in M17 using the
XIS onboard Suzaku. High signal-to-noise ratio spectra of the diffuse emission were
obtained in a spatially-resolved manner. Half a dozen of emission lines were resolved
clearly for the first time, which include K$\alpha$ lines from highly ionized O, Ne, and
Mg and L lines from Fe. These lines are clear evidence for the thermal origin of the
diffuse emission.

Based on the Chandra data of an overlapping field and on the previous works in the
literature, we confirmed that the background-subtracted diffuse spectrum by XIS is
barely contaminated by unresolved point sources, GRXE, CXB, and LHB emission.

We showed that the diffuse spectra are explained by a single temperature plasma model of
$k_{\rm{B}}T \sim$~0.25~keV, $L_{\rm{X}} \sim$~3.5$\times$10$^{33}$~erg~s$^{-1}$
(0.5--2.0~keV), and $N_{\rm H} \sim$4.5$\times$10$^{21}$~cm$^{-2}$. The temperature and
the chemical composition of the diffuse plasma is spatially uniform, indicating that the
plasma is at a thermal equilibrium. The apparent difference in the surface brightness is
probably due to the difference in the line-of-sight depth or in the filling factors, and
not in the plasma density.

The abundance is obtained individually for the detected elements, which are consistent
with 0.1--0.3~solar values. The enhancement of Ne against other metals by a factor of
$\sim$2 is seen in the diffuse plasma, which may be explained by a upward revision of
the solar Ne abundance.

The lack of spatial variations in the chemical composition comprises a sharp contrast to
the diffuse emission in the Carina Nebula, where different abundance patterns across a
similar spatial scale suggest the supernova origin for the emission. Together with the O
to Fe ratio consistent with the solar value, this gives evidence against the
interpretation that a supernova is the cause of the diffuse emission observed in M17.

\bigskip

The authors thank Takashi Hosokawa for useful discussion. Y.\,H. and M.\,T. acknowledge
financial support from the Japan Society for the Promotion of Science. The work is
supported by the Grants-in-Aid for the 21st century center of excellence program
``Center for Diversity and Universality in Physics'' and for the program number 18204015
from the Ministry of Education, Culture, Sports, Science and Technology of Japan. The
research made use of data obtained from the Data ARchive and Transmission System
(DARTS), provided by the PLAIN center, ISAS/JAXA.


\end{document}